\documentclass{article}
\usepackage{amsfonts,amssymb, amsmath,mathrsfs}

\usepackage[english]{babel}

\def\nn{\nonumber }
\def\bq{ \begin{equation} }
\def\eq{ \end{equation} }
\def\ben{ \begin{eqnarray} }
\def\en{ \end{eqnarray} }

\def\g{{\gamma}}

\newtheorem{prop}{Proposition}

\begin{document}


\title{On the Routh sphere problem}

\author{I.A. Bizyaev, A.V. Tsiganov \\
\it\small
Institute of Computer Science, Udmurt State University, Russia\\
\it\small e--mail:  bizaev\_90@mail.ru\\
\it\small St.Petersburg State University, St.Petersburg, Russia\\
\it\small e--mail:  andrey.tsiganov@gmail.com}

\date{}
\maketitle

\begin{abstract}
We discuss an embedding of a vector field for the nonholonomic Routh sphere into a subgroup of commuting Hamiltonian vector fields on  six dimensional phase space.  The corresponding Poisson brackets  are reduced to the canonical Poisson brackets on the Lie algebra  $e^*(3)$.  It allows us to relate nonholonomic Routh system with the Hamiltonian system on cotangent bundle to the sphere with  canonical Poisson structure.
 \end{abstract}

\section{Introduction}
\setcounter{equation}{0}

Let us consider a smooth  manifold $\mathcal M$ with coordinates  $x_1,\ldots x_m$ and a dynamical system defined by the following equations of motion
\bq\label{d-eq}
\dot{x}_i=X_i\,,\qquad i=1,\ldots,m.
\eq
We can identify this system of ODE's  with  vector field
\bq\label{d-vp}
X=\sum_{i=1}^m X_i\dfrac{\partial}{\partial x_i}\,,
\eq
which is a linear operator on a space of the smooth functions on $\mathcal M$
that encodes the infinitesimal evolution of any quantity
\[
\dot{F}=X(F)=\sum X_i \dfrac{\partial F}{\partial x_i}
\]
along the solutions of the system of equations  (\ref{d-eq}).

 In Hamiltonian mechanics  a Hamilton function $H$ on $\mathcal M$ generates vector field $X$ describing a dynamical system
\bq\label{ham-vp}
X=X_H\equiv PdH\,.
\eq
Here $dH$ is a  differential of   $H$ and  $P$ is some bivector on the phase space $\mathcal M$.
By adding some other assumptions we can prove that $P$ is  a Poisson bivector. In fact  it is enough to add  energy conservation
\[\dot{H}=X_H(H)=(PdH,dH)=0\]
and compatibility of dynamical evolutions associated with two functions $H_{1,2}$
\[
X_{H_1}(X_{H_2}(F))=X_{H_2}(X_{H_1}(F))+X_{X_{H_1}(H_2)}(F)\,,
\]
see  \cite{am78,jo64} and references within.

 For a considerable collection of nonholonomic dynamical systems vector fields are created using Hamilton function $H$  and density $g$  of  the invariant with respect to  $X$ measure
 \bq\label{cham-vp}
 X=gPdH\,.
 \eq
 This vector field  $X$ (\ref{cham-vp})  is a so-called conformally  Hamiltonian vector field,
 see examples of such fields in  \cite{bm01,bm02,bmk02, bbm11,ts12,ts12b}.

 Below we discuss nonholonomic Routh sphere problem with the vector field  $X$, which is a sum of commuting vector fields $PdH_k$ determined by  integrals of motion   $H_1,\ldots, H_n$:
\bq\label{ob-vp}
 X=g_1PdH_1+\cdots+g_n PdH_n\,.
 \eq
 Such decompositions are well-known in bi-Hamiltonian geometry \cite{ts11}  and for some other non\-ho\-lo\-no\-mic dynamical systems  \cite{bkm11,ts12}. If we have decomposition (\ref{ob-vp}) we can say that common levels of integrals of motion form a Lagrangian foliation associated with the Poisson bivector $P$ with all the ensuing consequences. We want to highlight that we discuss only properties of foliations and do not discuss linearization of the corresponding flows.

 It is known that a proper momentum mapping for the  non-Hamiltonian vector field  $X$ associated with the  Routh sphere  has a "focus-focus" singularity   \cite{cus98}. According to  \cite{d80} the nontriviality of the corresponding monodromy is a coarest obstruction to the existance of global action-angle variables.

\section{The Routh sphere}
\setcounter{equation}{0}
Following to \cite{bm02,ch48,cus98,rou55} let us consider  a rolling of dynamically symmetric and non balanced spherical rigid body, the so-called Routh sphere,  over a horizontal plane without slipping.
Dynamically non balanced means that that the geometric center differs on the center of mass, whereas dynamically symmetric means that two momenta of inertia coincide to each other, for instance  $I_1=I_2$. The line joining the center of mass and the geometric center is axe of inertial symmetry.

The moving sphere is subject to two kinds of constraint: a holonomic constraint of moving over of a horisontal plane and no slip nonholonomic constraint  associated with the zero velocity in the point of contact
\bq\label{rel-ch}
v+\omega\times r=0.
\eq
Here  $\omega$ and $v$ are the angular velocity and velocity of the center of mass of the ball,
$r$ is the vector joining the center of mass with the contact point and  $\times$ means the vector product in $\mathbb R^3$. All the vectors are expressed in the so-called body frame,  which is firmly attached to the ball, its origin is located at the center of mass of the body, and its axes coincide with the principal inertia axes of the body.

In the body frame   the angular momentum  $M$ of the ball with respect to the contact point is equal to
\bq\label{om-m}
M=\mathbf{I_Q} \omega\,,\qquad \mathbf{I_Q}=\mathbf{I}+mr^2\mathbf E-mr\otimes r.
\eq
Here $\mathbf E$ is a unit matrix, $m$ is a mass and $\mathbf I = \mathrm{diag}(I_1, I_2, I_3 )$  is an inertia tensor  of the rolling ball, respectively.

If  $\g=(\g_1,\g_2,\g_3)$  is the unit normal vector to the plane at the contact point, then
\[ r=(R\g_1,R\g_2, R\g_3+a)\,,\]
where $R$ is a radius of the ball and $a$ is a distance from geometric center to the center of mass.

The phase space, initial equations of motion, reduction of symmetries  and deriving of the reduced equation of motion are discussed in  \cite{bm02, bmk02, bm08,ch48, cus98}. We omit this step and begin directly  with the  reduced equations of motion on the six dimensional phase space $\mathcal M$ with local coordinates  $x=(\g,M)$ :
\bq\label{m-eq}\dot M=M\times \omega+m\dot{r}\times(\omega\times r)\,,\qquad \dot \g=\g\times \omega\,.
\eq
A straightforward calculation shows that these equations  (\ref{m-eq}) possess four integrals of motion
\bq\label{int}
H_1=(M,\omega)\,,\qquad H_2=(M,M)-mr^2H_1\,,\qquad H_3=(M,r)\qquad  H_4=(\g,\g)\,,
\eq
and the following invariant measure
\bq\label{rho}
\quad \mu= {g^{-1}(\g)\,}\,\mathrm d\g \,\mathrm dM\,,\qquad
g(\g)=\sqrt{I_1I_3+I_1mR^2(H_4-\g_3^2)+I_3m(R\g_3+a)^2}\,.
\eq
As for the symmetry Lagrange top there are two linear in momenta integrals of motion. First integral
\[H_3=(M,r)\]
  is a well-known Jellet integral \cite{jell72},  see also  \S243,  p. 192 in the Routh book \cite{rou55} .
Second integral
\bq\label{raus-int}
\hat{H}_2=g(\g)\,\omega_3\,,
\eq
was found by Routh in  1884  \cite{rou55} and recovered later by Chaplygin in \cite{ch48}.
These linear integrals are related with quadratic in momenta integral of motion
\[
H_2=I_1H_1+\dfrac{I_3-I_1}{I_1}\,\hat{H}_2^2-\dfrac{m}{I_1}\,H_3^2\,.
\]
According to  \cite{bm02} namely this quadratic integral survives at $I_1\neq I_2$ in contrast with the linear integral $\hat{H}_2$.

If the Routh sphere rolls on a horizontal plane under the influence of a constant vertical gravitational force, equations of motion read as
\bq\label{mg-eq}
\dot M=M\times \omega+m\dot{r}\times(\omega\times r)+\g \times \frac{\partial U}{\partial \gamma}\,,\qquad \dot \g=\g\times \omega\,.
\eq
Here  $U=-m\mathrm g(r,\g)$ and  $\mathrm g$ is a gravitational acceleration.  These equations possess the same invariant measure $\mu$  and integrals  $H_3$, $H_4$,$\hat{H}_2$, whereas quadratic integrals of motion  are equal to
\bq\label{int-g}
{H}_1=(\omega,M)+2U\,,\qquad   {H}_2=M^2-mr^2(\omega,M)-2m\mathrm g\Bigl(a I_1\gamma_3-mR^2\gamma^2(a^2+\gamma^2R^2)\Bigr).
\eq
In this case vector field $X$ defined by equations (\ref{mg-eq}) has homoclinic trajectories when the Routh sphere is either spinning very slowly about a vertical axes which passes through the center of mass and the geometric center, either axis is in the same position but the value of the Jellet integral  is slightly less than the threshold value \cite{cus98}.

\subsection{Poisson brackets}
For the Routh sphere six equations of motion  (\ref{m-eq})  possess four integrals of motion and an invariant measure and, therefore,  they are integrable by quadratures according to the  Euler-Jacobi theorem. It allows us to suppose that
 common level surfaces of integrals form  a direct sum of  symplectic and lagrangian foliations of
 dual dynamical system which is hamiltonian with respect to the Poisson bivector $P$, so  that

\bq\label{geom-eq}
 [P,P]=0\,,\qquad PdC_{1,2}\equiv P\mathrm dH_{i,j}=0\,,\qquad (P\mathrm dH_\ell,\mathrm dH_k)\equiv\{H_\ell,H_k\}=0\,.
\eq
Here $[.,.]$ is the  Schouten bracket and $(i,j,l,m)$ is the arbitrary permutation of $(1,2,3,4)$. In fact, here we suppose that the Euler-Jacobi  integrability of non-Hamiltonian system (\ref{m-eq}) is equivalent to the Liouville integrability of the
 dual Hamiltonian dynamical system with the same integrals of motion, see \cite{ts11}.

The first equation in (\ref{geom-eq}) guaranties that $P$ is a Poisson bivector. In the second equation  we define  two Casimir elements $\mathcal H_{i}$ and $\mathcal H_j$ of $P$ and assume that rank$P=4$. It is a necessary condition because by fixing its values one gets the four dimensional symplectic phase space of our dynamical system. The third equation provides that the two remaining integrals $\mathcal H_\ell$ and $\mathcal H_k$ are in involution with respect to the Poisson bracket associated with $P$.

In this note we discuss solutions of the equations (\ref{geom-eq}) in the space of the linear in momenta  $M_i$ bivectors $P$ at the  different choice of the  Casimir functions:
\begin{enumerate}
                     \item $C_1=(\g,\g)\,,\qquad C_2=(M,r)\,,\qquad H_\ell=H_1\,,\qquad H_k=H_2 $;
                     \item $C_1=(\g,\g)\,,  \qquad C_2=g(\g)\,\omega_3\qquad H_\ell=H_1\,,\qquad H_k=H_3 $;
                     \item $C_1=(M,r)\,, \qquad C_2=g(\g)\,\omega_3\qquad H_\ell=H_1\,,\qquad H_k=H_4 $.
\end{enumerate}
In generic case  linear in momenta Casimir functions look like
\bq\label{gen-caz}
C_{1,2}=a_{1,2}(\g,\g)+b_{1,2}(M,r)+c_{1,2}g(\g)\,\omega_3\,,\qquad a_{1,2},b_{1,2},c_{1,2}\in\mathbb C\,.
\eq
However, the corresponding complete solutions of (\ref{geom-eq}) have the same properties as particular solutions obtained in the  listed above three special cases.

In \cite{ts08l} we have solved the same system of equations  (\ref{geom-eq}) for the symmetric Lagrange top and proved that solutions may be useful for construction of the variables of separation and the recursion Lenard-Magri relations for this Hamiltonian system.

If we have some solution $P$ of  (\ref{geom-eq}), we can get decomposition of initial vector field   $X$ by commuting Hamiltonian vector fields  $PdH_\ell$ and $PdH_k$. The existence of such decomposition  by the basis of the Hamiltonian vector fields requires to impose one more condition  rank$P$=4.

\section{First  Poisson bracket}
\setcounter{equation}{0}
Substituting linear in momenta  $M_i$ anzats for entries of the Poisson bivector
\bq\label{anzats}
P_{ij}=\sum_{k=1}^3 a_{ijk}(\g)M_k+b_{ij}(\g)
\eq
into   (\ref{geom-eq}) at
\[C_1=H_4=(\g,\g)\,,\qquad C_2=H_3=(M,r)\,,\qquad H_\ell=H_1\,,\qquad H_k=H_2\]
and solving the resulting system of algebro-differential equations one gets the following Proposition.
\begin{prop}
In this case generic solution of  (\ref{geom-eq}) is parameterized by two functions  $\alpha(\g_1/\g_2)$ and $\beta(\g_3)$
\bq\label{p-bv}
P=\alpha g \left(
                       \begin{array}{cc}
                         0 & \mathbf\Gamma_\alpha \\
                         -\mathbf\Gamma^\top_\alpha & \mathbf M_\alpha \\
                       \end{array}
                     \right)+\beta\left(
                       \begin{array}{cc}
                         0 & \mathbf\Gamma_\beta \\
                        - \mathbf\Gamma^\top_\beta & \mathbf M_\beta \\
                       \end{array}
                     \right)
\,.\eq
Here matrices $\Gamma_{\alpha,\beta}$ are equal to
\[
\mathbf\Gamma_\alpha=\left(
                \begin{array}{ccc}
                  \frac{\g_1\g_2(R\g_3+a)}{R(\g_1^2+\g_2^2)} & \frac{\g_2^2(R\g_3+a)}{R(\g_1^2+\g_2^2)} & -\g_2 \\
                  -\frac{\g_1^2(R\g_3+a)}{R(\g_1^2+\g_2^2)} & -\frac{\g_1\g_2(R\g_3+a)}{R(\g_1^2+\g_2^2)} & \g_1 \\
                  0 & 0 & 0 \\
                \end{array}
              \right)\,,\qquad
\mathbf\Gamma_\beta=\left(
               \begin{array}{ccc}
                 --\frac{\g_1\g_2\g_3}{\g_1^2+\g_2^2} & \frac{\g_1^2\g_3}{\g_1^2+\g_2^2}  & 0 \\
                 -\frac{\g_2^2\g_3}{\g_1^2+\g_2^2}  & \frac{\g_1\g_2\g_3}{\g_1^2+\g_2^2}  & 0 \\
                 \g_2 & -\g_1 & 0 \\
               \end{array}
             \right)\,,
\]
and skew symmetric  matrices $M_{\alpha,\beta}$ have the form
\ben
\mathbf M_\alpha&=&\left(
           \begin{array}{ccc}
             0 & \dfrac{(\gamma_1M_1+\gamma_2M_2)(R\gamma_3+a)}{R(\g_1^2+\g_2^2)} & -M_2 \\
             * & 0 & M_1 \\
             * & * & 0 \\
           \end{array}
         \right)\,,\nn\\
         \nn\\
\mathbf M_\beta&=&\left(
          \begin{array}{ccc}
            0 & M_3-\dfrac{\gamma_3(\gamma_1M_1+\gamma_2M_2)}{\gamma_1^2+\gamma_2^2}-\dfrac{m\sigma R(R\gamma_3+a)}{g^2} & \dfrac{m\sigma \gamma_2R^2}{g^2} \\
            * & 0 &  -\dfrac{m\sigma \gamma_1R^2}{g^2} \\
            * & * & 0 \\
          \end{array}
        \right)\,,\nn
\en
where $g\equiv g(\g)$ and
\[
\sigma=mR\Bigl(m(r,\g)C_2+I_3(\g_1M_1+\g_2M_2)+I_1M_3\g_3\Bigr)\,.
\]
\end{prop}
The proof is a straightforward solution of  (\ref{geom-eq}) using linear in momenta anzats.

The corresponding Poisson brackets read as
\bq\label{p-br}
\begin{array}{ll}
\{M_1,\g_1\}=-\dfrac{\alpha g \g_1\g_2(R\g_3+a)}{(\g_1^2+\g_2^2)R}+\dfrac{\beta\g_1\g_2\g_3}{\g_1^2+\g_2^2}\,,\qquad&
\{M_1,\g_2\}=\dfrac{\alpha g \g_1^2(R\g_3+a)}{(\g_1^2+\g_2^2)R}+\dfrac{\beta\g_2^2\g_3}{\g_1^2+\g_2^2}\,,
\\
\\
\{M_1,\g_3\}=-\beta\g_2\,,& \{M_2,\g_3\}=\beta\g_1\,,\\
\\
\{M_2,\g_1\}=-\dfrac{\alpha g \g_2^2(R\g_3+a)}{(\g_1^2+\g_2^2)R}-\dfrac{\beta \g_1^2\g_3}{\g_1^2+\g_2^2}\,,\qquad&
\{M_2,\g_2\}=\dfrac{\alpha g \g_1\g_2(R\g_3+a)}{(\g_1^2+\g_2^2)R}-\dfrac{\beta \g_1\g_2\g_3}{\g_1^2+\g_2^2}\,,\\
\\
\{M_3,\g_1\}=\alpha g \g_2\,,\qquad \{M_3,\g_2\}=-\alpha g \g_1\,,\qquad & \{M_3,\g_3\}=0\,,\quad \{\g_i,\g_j\}=0\,,
\end{array}
\eq
and
\[
\{M_1,M_2\}=\dfrac{\alpha g (\g_1M_1+\g_2M_2)(R\g_3+a)}{(\g_1^2+\g_2^2)R}+
\beta\left(
M_3-\dfrac{\g_3(\g_1M_1+\g_2M_2)}{\g_1^2+\g_2^2}-\dfrac{\sigma\,(R\g_3+a)}{g^2}
\right)\,,
\]
\[
\{M_1,M_3\}=-\alpha g  M_2+\dfrac{\beta\sigma R}{g^2}\,\g_2\,,\qquad
\{M_2,M_3\}=\alpha g M_1-\dfrac{\beta\sigma R}{g^2}\, \g_1\,.
\]
  In generic case {rank}$P=4$, however if $\alpha=0$  or  $\beta=0$ there are additional Casimir function $\g_3$ and $\g_1/\g_2$, respectively. A particular form of this brackets was obtained in  \cite{ram}.

Using this Poisson bivector we can obtain  basis of the commuting Hamiltonian vector fields \[X_1=PdH_1\qquad\mbox{and}\qquad X_2=PdH_2\,,\]
and try to expand initial non-Hamiltonian vector field $X$
 (\ref{mg-eq}) by these vector fields.
\begin{prop}
Using Poisson brackets  (\ref{p-br}) we can rewrite reduced equations of motion for the Routh sphere  (\ref{mg-eq}) in the following form
\bq\label{g-eq1}
\dot{x}_k=g_1\{x_k,H_1\}+g_2\{x_k,H_2\},\qquad k=1,..6\,,
\eq
if and only if
\bq\label{cond}
\alpha(\g_1/\g_2)=\mathrm{const}\,, \qquad\beta(\g_3) = \alpha g  \left(1+\dfrac{a}{R\g_3}\right)\,.
\eq
In this case  coefficients  are equal to
\[
g_1=-\dfrac{(R\g_3+a)I_1-R\g_3
I_3}{2\alpha g (I_1-I_3)(R\g_3+a)}\,,\qquad g_2=\dfrac{a}{2\alpha g (I_1-I_3)(R\g_3+a)}\,.
\]
\end{prop}
The proof is a straightforward verification of the equations (\ref{g-eq1}).

There are other special values of the functions  $\alpha$ and $\beta$ according to the following
\begin{prop}
If
\bq\label{cond2}
\alpha(\g_1/\g_2)=\mathrm{const}\,, \qquad\mbox{and}\qquad \beta(\g_3)= \dfrac{\alpha g (I_1R\g_3-I_3(R\g_3+a)}{R\bigl((I_1-I_3)\g_3+am(r,\g)\bigr)}\,,
\eq
the Poisson bivector $P$ (\ref{p-bv}) is compatible with the canonical Poisson
bivector $P_0$ on the Lie algebra $e^*(3)$
\[
P_0=\left(\begin{array}{cc}0&\mathbf \Gamma\\ -\mathbf \Gamma^\top&
\mathbf M\end{array}\right)\,,
\]
where
\[
\mathbf \Gamma=\left(
                 \begin{array}{ccc}
                   0 & \g_3 & -\g_2 \\
                   -\g_3 & 0 & \g_1 \\
                   \g_2 & -\g_1 & 0
                 \end{array}
               \right)\,,\qquad
\mathbf M=\left(
                 \begin{array}{ccc}
                   0 & M_3 & -M_2 \\
                   -M_3 & 0 & M_1 \\
                   M_2 & -M_1 & 0
                 \end{array}
               \right)\,.
\]
\end{prop}
The proof is a calculation of the Schouten bracket $[P_0,P]=0$.

Remind, compatibility means that a linear combination of this bivectors
\[ P_\lambda=P_0+\lambda P\,,\qquad \lambda\in \mathbf C\,,\]
is a Poisson bivector at any value of   $\lambda$. It also means that $P$ (\ref{p-bv}) is a trivial deformationof $P_0$, see details in  \cite{ts11ch}.

It is interesting that namely  this condition of compatibility allows us to expand initial vector field  (\ref{mg-eq})  by a basis of Hamiltonian vector fields
\bq\label{g-eq2}
\dot{x}_k=\hat{g}_1\{x_k,H_1\}+\hat{g}_2\{x_k,\hat{H}_2\},\qquad k=1,..6\,,
\eq
associated with the linear in momenta Routh integral $\hat{H}_2$   (\ref{raus-int}). Here  coefficients
\[
\hat{g}_1=-\dfrac{1}{2\beta}\,,\qquad\qquad \hat{g}_2=\dfrac{a}{\alpha I_1\bigl(I_1R\g_3-I_3(R\g_3+a)\bigr)}\left(
M_3-\dfrac{\g_3(\g_1M_1+\g_2M_2)}{\g_1^2+\g_2^2}
\right)
\]
depend on coordinates and momenta in contrast with the previous decomposition.

Summing up, for the Routh sphere integrals of motion  are in involution at any values of  $\alpha$ and $\beta$
\[\{H_1,H_2\}=\{H_1,\hat{H}_2\}=0\,,\qquad \forall \alpha,\beta\,.\]
However, initial vector field $X$ is decomposed by the corresponding commuting Hamiltonian vector fields
\[
X=g_1PdH_1+g_2PdH_2\,,\quad\mbox{or}\quad X=\hat{g}_1PdH_1+\hat{g}_2Pd\hat{H}_2
\]
only at the special values of these functions (\ref{cond}) or (\ref{cond2}),  respectively.

\subsection{Properties of the first Poisson brackets}
Similar to Chaplygin sphere problem \cite{ts12} and to nonholonomic Veselova problem  \cite{ts12b} we can reduce this Poisson bracket to the canonical Poisson brackets on the Lie algebra  $e^*(3)$ and identify the Routh sphere model with
the Hamiltonian system on  two-dimensional sphere.

One of the possible reductions is given by the following Proposition.
\begin{prop}
After a change of momenta
\bq\label{p-map}
\begin{array}{l}
L_1=\frac{1}{\g_1^2+\g_2^2}\left(\frac{\g_1\g_3\left(R(\g_1M_1+\g_2M_2)-bI_1^{-1}(I_1+m(R\g_3+a)^2\right)}{\alpha g(R\g_3+a)}
+\frac{\g_2(\g_2M_1-\g_1M_2)}{\beta}+c\g_1
\right)\,,\\
\\
L_2=\frac{1}{\g_1^2+\g_2^2}\left(\frac{\g_2\g_3\left(R(\g_1M_1+\g_2M_2)-bI_1^{-1}(I_1+m(R\g_3+a)^2\right)}{\alpha g(R\g_3+a)}
-\frac{\g_1(\g_2M_1-\g_1M_2)}{\beta}+c\g_2
\right)\,,\\
\\
L_3 =\frac{ M_3}{\alpha g}+\frac{bm(R\g_3+a)}{\alpha g I_1}\,,\qquad \qquad b=(M,r)\,,\quad c=(L,\g)\,,
\end{array}
\eq
the Poisson brackets $\{.,.\}$ (\ref{p-br}) coincide with the canonical Poisson brackets on the Lie algebra
 $e^*(3)$
\begin{equation}\label{e3}
\bigl\{L_i\,,L_j\,\bigr\}_0=\varepsilon_{ijk}L_k\,,
 \qquad
\bigl\{L_i\,,\g_j\,\bigr\}_0=\varepsilon_{ijk}\g_k \,,
\qquad
\bigl\{\g_i\,,\g_j\,\bigr\}_0=0\,,
\end{equation}
where $\varepsilon_{ijk}$ is a completely antisymmetric tensor.
\end{prop}
It is easy to see that the Poisson map  (\ref{p-map}) is  locally defined in the region \[\gamma_1^2+\gamma_2^2\equiv1-\gamma_3^2\neq 0.\]
Namely in this region of the phase space the vector field  for the Routh sphere  $X$ (\ref{mg-eq}) does not have homoclinic orbits \cite{cus98}.

If $c=(\g,L)=0$ and condition  (\ref{cond}) holds, then images of the initial integrals of motion are nonhomogeneous second order polynomials in momenta
\ben
{H}_1&=&\frac{\alpha^2}{\gamma_1^2+\gamma_2^2}\left(\frac{g^2-I_1I_3}{mR^2}L_3^2+\frac{g^2(R\gamma_3
+a)^2(L_1\gamma_2-L_2\gamma_1)^2}{R^2\gamma_3^2(I_1+mr^2)} \right)\nn\\
\nn\\
&-&\dfrac{2b\alpha g (R\g_3+a)L_3}{I_1R^2(\g_1^2+\g_2^2)}
+\dfrac{b^2(I_1+m(R\g_3+a)^2)}{I_1^2R^2(\g_1^2+\g_2^2)}\,,\nn\\
\label{int-sph}\\
{H}_2&=&\frac{\alpha^2}{\gamma_1^2+\gamma_2^2}\left(\frac{I_1I_3r^2}{R^2}L_3^2+\frac{g^2I_1(R\gamma_3
+a)^2(L_1\gamma_2-L_2\gamma_1)^2}{R^2\gamma_3^2(I_1+mr^2)} \right)\nn\\
\nn\\
&-&\dfrac{2b\alpha g(R\g_3+a)L_3}{R^2(\g_1^2+\g_2^2)}
+\left(\dfrac{I_1+mr^2}{I_1R^2(\gamma_1^2+\gamma_2^2)}+\dfrac{2m}{I_1}\right)b^2.\nn
\en
These integrals define a Hamiltonian system on cotangent bundle $T^*S^2$ to the  sphere $S^2$.

As for Lagrange top the existence of the linear in momenta integral of motion  (\ref{raus-int})
\[
\hat{H}_2=\alpha I_1 L_3\,,
\]
allows us to explicitly integrate the corresponding Hamiltonian equations of motion by quadratures.  Namely, let us introduce spherical coordinates on the sphere
\bq\label{sph-coord}
\begin{array}{ll}
\g_1 =\sin\phi\sin\theta,\qquad &
L_1 =\dfrac{\sin\phi\cos\theta}{\sin\theta}\,p_\phi-\cos\phi\,p_\theta\,,\\
\\
\g_2 = \cos\phi\sin\theta,\qquad&
L_2=\dfrac{\cos\phi\cos\theta}{\sin\theta}\,p_\phi+\sin\phi\,p_\theta\,, \\
\\
\g_3 =\cos\theta\,,\qquad &L_3 = -{p_\phi}\,,
\end{array}
\eq
where $\phi,\theta$ are the Euler angles,  $p_\phi$ and $p_\theta$ are the canonically conjugated momenta
\[
\{\phi,p_\phi\}=\{\theta,p_\theta\}=1\,,\qquad \{\phi,\theta\}=\{\phi,p_\theta\}=\{\theta,p_\phi\}=0\,.\]
In this variables initial integrals of motion are equal to
\bq\label{int-sph2}
{H}_1=A(\theta)\,p_\phi^2+B(\theta)\,p_\theta^2+bC(\theta)\,p_\phi+b^2V(\theta)\,,\qquad \hat{H}_2=-\alpha I_1\,p_\phi\,,
\eq
where $b$  is a value of the Jellet integral,  $A,B,C$ and $V$  are function on  $\theta$:
\[\begin{array}{ll}
A(\theta)=\alpha^2\left(I_1+\frac{I_3(a^2+2aR\cos\theta+R^2\cos^2\theta)}{R^2\sin^2\theta}\right)\,,\quad&
B(\theta)=\frac{\beta^2}{I_1+m(a^2+2aR\cos\theta+R^2)}\,,\\
\\
C(\theta)=\frac{2\alpha g(R\cos\theta+a)}{I_1R^2\sin^2\theta}\,,\qquad &
V(\theta)=\frac{I_1+m(a^2+2aR\cos\theta+R^2)}{I_1^2R^2\sin^2\theta}\,.
\end{array}
\]
Thus, similar to the Lagrange top, solution of the Hamiltonian equations of motion generated by the Hamilton function and canonical Poisson brackets is reduced to solution of one equation
\[
\dot{\theta}=2B(\theta)\,p_\theta=2\sqrt{B(\theta)\bigl(E-A(\theta)c^2-bcC(\theta)-b^2 V(\theta)\bigr)}\,,
\]
where $E={H}_1$ and $c=-\hat{H}_2/\alpha I_1$  are constants of motion.

Summing up, after reduction of the Poisson brackets to canonical ones we can easily find trajectories of the commuting Hamiltonian vector fields
\[X_1=PdH_1,\qquad X_2=PdH_2 \quad\mbox{and}\quad \hat{X}_2=Pd\hat{H}_2\,.\]
However, in order to get solutions of the Routh equations of motion we have to integrate their linear combinations
\[X=g_1X_1+g_2X_2\qquad\mbox{or}\qquad X=\hat{g}_1X_1+\hat{g}_2\hat{X}_2\,,\]
associated with conditions (\ref{g-eq1}) or (\ref{g-eq1}), respectively. Solution of this problem remains open as of yet.

\subsection{Conformally Hamiltonian equations of motion}
 If the Jellet integral is equal to zero  $C_2=(M,r)=0$, i.e. if  $b=0$, then integrals of motion $H_{1,2}$ (\ref{int-sph}) become homogeneous quadratic polynomials in momenta. In this case we can easily find variables of separation  $q_{1,2}$ in the corresponding Hamilton-Jacobi equation, if we  diagonalize simultaneously  two quadratic forms   $H_{1,2}$ (\ref{int-sph}).  Then,  using  these variables of separation, we can rewrite initial vector field $X$  (\ref{ob-vp})  in the conformally Hamiltonian form (\ref{cham-vp}).

Namely,  at $C_2=b=0$ integrals of motion  $H_{1,2}$ satisfy to the following separated relations
 \[
 \Phi_i(q_i,p_{i},H_1,H_2)=0\,,\qquad k=1,2.
\]
Here $q_{1,2}$ and $p_{1,2}$ are canonically conjugated variables of separation. In this case, according to  \cite{ts07} ,  these integrals  $H_{1,2}$ are in involution
 \[\{H_1,H_2\}_f=0\]
 with respect to the Poisson brackets
 \[
\{q_1,p_{1}\}_f=f_1(q_1,p_{1})\,,\qquad \{q_2,p_{2}\}_f=f_2(q_2,p_{2})\,,
\qquad \{q_1,q_2\}_f=\{p_{1},p_{2}\}_f=0\,,
\]
labelled by two arbitrary functions $f_{1,2}$.  The corresponding Poisson bivector $P_f$ and integrals of motion  $H_{1,2}$
satisfy to the equations
\bq\label{f-mat}
P_fdH_i=F_{i1}\,PdH_1+F_{i2}\,PdH_2\,,\qquad i=1,2.
\eq
Here functions $F_{ij}$ depend on  $f_{1,2}$ and form the so-called control matrix \cite{ts07,ts11}.
\begin{prop}
If $X$ (\ref{ob-vp}) is a linear combination of the commuting Hamiltonian vector fields
\[
 X=g_1PdH_1+g_2 PdH_2\,,
 \]
 and coefficients $g_{1,2}$ are special combinations of   $F_{ij}$
 \[
 g_i=g \Bigl(a_1 F_{i1}+a_2F_{i2}\Bigr)\,,\qquad i=1,2,
 \]
there is a Poisson bivector  $P_f$, which allows us to rewrite  $X$ in the conformally Hamiltonian form
 \[
 X=g_1PdH_1+g_2 PdH_2=g P_f d H\,,\qquad  H=a_1H_1+a_2H_2.
 \]
 In this case $H$ is  a sum of initial physical integrals of motion  $H_{1,2}$.
\end{prop}
For the Routh sphere at $C_2=0$ variables of separation $q_{1,2}$ are functions only  on  coordinates $\g_i$. Thus, the desired bivector  $P_f$ may be directly  obtained from  $P$ (\ref{p-bv}) at
\[
\alpha=-R\,,\qquad\beta=-\,\dfrac{g\,(r,\mathbf I r)}{(\g,\mathbf I r)}\,,
\]
so we have
\[
 X=g_1PdH_1+g_2 PdH_2=-\dfrac{1}{2\beta}\, P_f dH_1\,.
\]
At  $C_2=0$ variables of separation  $q_{1,2}$  have to be functions on coordinates  $\g_i$ and  momenta  $M_i$ and, therefore,  entries of  $P_f$  have to be  more complicated  functions on $M_i$. Unfortunately, we do not know how to get variables of separation for the nonhomogeneous polynomial integrals of motion  (\ref{int-sph}) on the sphere.

\section{Second and third Poisson brackets}
\setcounter{equation}{0}
Let us substitute  linear in momenta  $M_i$ anzats (\ref{anzats}) into the equations  (\ref{geom-eq}) at
\[C_1=H_4=(\g,\g)\,,\qquad C_2=\hat{H}_2=g(\g)\,\omega_3\,,\qquad H_\ell=H_1\,,\qquad H_k=H_3\,.\]
\begin{prop}
In this case  integrals of motion are in involution $\{H_\ell,H_k\}=0$  if and only if   bivector
\bq\label{p-bv2}
P'= \left(
                       \begin{array}{cc}
                         0 & \mathbf\Gamma' \\
                         -\mathbf\Gamma'^\top & \mathbf M'\\
                       \end{array}
                     \right)
\,,\eq
 is labelled by three arbitrary functions  $\alpha_i(\g)$ entering into the  matrix
\[
\mathbf\Gamma'=\left(
                  \begin{array}{ccc}
                   -\g_2(\alpha_1+\g_3\alpha_3)&-\g_2 \alpha_2+\g_3 \g_1 \alpha_3  & \frac{m R \g_2 (\g_1 \alpha_1+\g_2 \alpha_2) (R \g_3+a)}{I_1+m(R\g_3+a)^2} \\
                   \g_1 \alpha_1& \g_1 \alpha_2 & -\frac{m R \g_1 (\g_1 \alpha_1+\g_2 \alpha_2) (R \g_3+a)}{I_1+m(R\g_3+a)^2} \\
                    \g_1 \g_2 \alpha_3 & -\g_1^2 \alpha_3 & 0 \\
                  \end{array}
                \right)\,,
\]
 and into the skew symmetric matrix
\ben
\mathbf M'_{1,2}&=&-\alpha_1M_1-\alpha_2M_2+\alpha_3(\g_1M_3-\g_3M_1)\nn\\
&-&\frac{\alpha_3\g_1R(R\g_3+a)\Bigl(m^2(\g,r)H_3+m\bigl(I_3(\g_1M_1+\g_2M_2)+I_1\g_3M_3\bigr)\Bigr)}{g^2}\,,\nn\\
\nn\\
\mathbf M'_{1,3}&=&\frac{mR(R\g_3+a)(\g_1\alpha_1+\g_2\alpha_2)}{I_1+m(R\g_3+a)^2}\,M_2\nn\\
&+&\frac{\alpha_3\g_1\g_2R^2\Bigl(m^2(\g,r)H_3+m\bigl(I_3(\g_1M_1+\g_2M_2)+I_1\g_3M_3\bigr)\Bigr)}{g^2}\,,\nn\\
\nn\\
\mathbf M'_{2,3}&=&-\frac{mR(R\g_3+a)(\g_1\alpha_1+\g_2\alpha_2)}{I_1+m(R\g_3+a)^2}\,M_1\nn\\
&-&\frac{\alpha_3\g_1^2R^2\Bigl(m^2(\g,r)H_3+m\bigl(I_3(\g_1M_1+\g_2M_2)+I_1\g_3M_3\bigr)\Bigr)}{g^2}\,.\nn
\en
If we  impose additional restriction  rank$P'=4$, the third equation $[P',P']=0$ in (\ref{geom-eq})  has the following single solution
\ben
\alpha_1& =& -\dfrac{I_1+m(R\g_3+a)^2}{\g_1}+
\dfrac{\g_2^2I_1R\bigl(I_1+ma(R\g_3+a)\bigr)}{\g_1(\mathbf I\g,r)}\,,\nn\\
\alpha_2&=&-\dfrac{ \g_2I_1R\bigl(I_1+ma(R\g_3+a)\bigr)}{(\mathbf I\g,r)}\\
\alpha_3&=&\dfrac{I_1+m(R\g_3+a)^2}{\g_1\g_3}-\dfrac{(\g_1^2+\g_2^2)RI_1\bigl(I_1+ma(R\g_3+a)\bigr)}{\g_1\g_3(\mathbf I\g,r)}\,.\nn
\en
\end{prop}
The proof is a straightforward solution of the differential equations using substitution (\ref{anzats}).

Using this Poisson bivector we can get a basis of Hamiltonian vector fields  and an expansion of the initial vector field
\bq\label{g-eq3}
X=g'_1P'dH_1+g'_2P'dH_3^2
\eq
by these vector fields. The corresponding coefficients are equal to
\[
g'_1=-\dfrac{1}{2\g_1\alpha_3 I_1}\,,\qquad g'_2=\dfrac{I_1(I_1+ma(R\g_3+a))-mR(\g,{\bf I}r)}{2g^2RI_1^2(R\g_3+a)}\,.
\]
Similar to the first Poisson bracket  (\ref{p-br}), there is  a transformation of momenta $M_i\to L_i$, which reduces this Poisson brackets $\{.,.\}'$ to canonical Poisson brackets on the Lie algebra ï $e^*(3)$ .

Now let us consider the third possible choice of the linear in momenta  Casimir functions
\[C_1=(M,r)\,, \qquad C_2=g(\g)\,\omega_3\qquad H_\ell=H_1\,,\qquad H_k=H_4 \,.\]
In this case generic solution of  (\ref{geom-eq}) coincides with the previous solution  $P'$ (\ref{p-bv2})
 \[P''=\left.P'\right|_{\g_1 \alpha_1+\g_2 \alpha_2=0}\,,\]
at $\g_1 \alpha_1+\g_2 \alpha_2=0$.  It is easy to see that rank$P''=3$ and there is a third Casimir function $H_4=(\g,\g)$
\[
P''dC_3=0\,,\qquad C_3=H_4\,.
\]
Consequently, in this case we have only one nontrivial Hamiltonian vector field  $P''dH_1$, which does not form a basis.

In the similar manner we can consider generic linear in momenta Casimir functions (\ref{gen-caz}).
\begin{prop}
We can not rewrite equations of motion  (\ref{mg-eq}) on the six-dimensional phase space
in the conformally Hamiltonian form
\bq\label{eq-cf}
 X=gPdF(H_1,H_2,H_3,H_4)
 \eq
using linear in momenta Poisson bivector $P$ satisfying equations
 (\ref{geom-eq}).  Here $F(H_1,H_2,H_3,H_4)$ is an arbitrary function on integrals of motion for the Routh sphere.
\end{prop}
The proof is a straightforward verification of the fact that common system of equations (\ref{eq-cf} and  (\ref{geom-eq}) is inconsistent at the space of the no more then  linear in momenta Casimir functions and Poisson bivectors.

\section{Conclusion}
It is well known that equations of motion for the nonholonomic Routh sphere are integrable by quadratures according to the Euler-Jacobi theorem.  We identify the corresponding  level sets of integrals of motion with the Lagrangian foliations associated with two different Poisson bivectors $P$ and $P'$.  The corresponding expansions of the initial vector field  $X$  (\ref{g-eq1}) and (\ref{g-eq3})
\[
X=g_1PdH_1+g_2PdH_2=g'_1P'dH_1+g'_2P'dH_3^2\,
\]
may be considered as  a counterpart of the standard Lenard-Magri recurrence relations
\[X=PdH_1=f_1P'dH_2+f_2P'dH_3\]
for two dimensional bi-Hamiltonian systems $(f_1=1,\,\quad f_2=0)$, quasi bi-Hamiltonian systems  $(f_2=0)$ or bi-integrable systems $(\forall f_{1,2})$, which appear in Hamiltonian mechanics \cite{ts07,ts08l,ts11}.

We would like to thank  A.V. Bolsinov, A.V. Borisov and I.S. Mamaev   for useful  discussion of applications of the Poisson geometry to the different  nonholonomic systems.

\end{document}